\documentclass[a4paper,11pt]{article} 

\usepackage{jcappub}

\usepackage[english]{babel}
\usepackage{amsfonts,amsmath,amssymb}
\usepackage{graphicx}
\usepackage{url}
\usepackage[normalem]{ulem} 
\usepackage{hyperref}

\newcommand{\lcdm}{$\Lambda$CDM\,}
\newcommand{\ar}{\alpha_{\mathrm{d}}}
\newcommand{\as}{\alpha_{\mathrm{c}}}

\newcommand{\eff}{\mathrm{eff}}
\newcommand{\enangle}[1]{\langle #1 \rangle} 
\newcommand{\mpl}{M_{\textrm{Pl}}}

\title{Dynamics of non-minimally coupled perfect fluids}

\author[a]{Dario Bettoni,}
\author[b,c]{Stefano Liberati%
}
\affiliation[a]{Faculty of Physics, Israel Institute of Technology,\\
Technion City, 32000, Haifa, Israel}
\affiliation[b]{SISSA,\\
Via Bonomea 265, 34136, Trieste, Italy}
\affiliation[c]{INFN, Sezione di Trieste, \\
Via Valerio, 2, 34127, Trieste, Italy}

\emailAdd{dario@physics.technion.ac.il}
\emailAdd{liberati@sissa.it}

\abstract{We present a general formulation of the theory for a non-minimally coupled perfect fluid in which both conformal and disformal couplings are present. We discuss how such non-minimal coupling is compatible with the assumptions of a perfect fluid and derive both the Einstein and the fluid equations for such model. We found that, while the Euler equation is significantly modified with the introduction of an extra force related to the local gradients of the curvature, the continuity equation is unaltered, thus allowing for the definition of conserved quantities along the fluid flow. As an application to cosmology and astrophysics we compute the effects of the non-minimal coupling on a Friedmann--Lema\^itre--Robertson--Walker metric at both background and linear perturbation level and on the Newtonian limit of our theory.}


\begin{document}

\maketitle
\flushbottom

\section{Introduction}

There is a general consensus that modern cosmology has entered a golden age, where increasingly accurate observational tools are allowing us to push our knowledge of the Universe beyond the limits of what was imaginable just a few decades ago. This grand project is resting on few theoretical pillars: The Copernican principle, the standard theory of gravitation i.e.~General Relativity (GR) and the big bang model of the Universe (including an inflationary phase). The combination of these operative frameworks has lead to a compelling coherent theory of the Universe which is often referred to as the concordance model or standard model of cosmology (\lcdm) which so far seem to explain very well current observations~\cite{Ade:2013zuv}. Nonetheless, in the last decade there has been a growing feeling in the cosmologist community that alternatives to or modification of this standard model are required.

There are several reasons one can adduce in support of this general trend, however, there is a general consensus that the main issue with the current paradigm is the requirement of an overwhelmingly dominating dark sector in the energy/matter composition of the universe (about 95\% of the total). Not only we do not have at the moment a definitive explanation for the origin of dark matter and dark energy but current observations are also questioning our capability of effectively describe these cosmological components with standard methods, e.g.~by modelling them on large scales as perfect fluids.

Indeed, approaches that attempt to describe Dark Energy (DE) as a fluid beyond the perfect fluid assumption \cite {Comer:1993zfa,Dubovsky:2011sj,Ballesteros:2012kv,Ballesteros:2014sxa} have been proposed, as well as fluid formulations of DE models based on scalar field  \cite{Pujolas:2011he,Sawicki:2012re} or assuming exotic equations of state \cite{Kamenshchik:2001cp,Bilic:2001cg,Bento:2002ps} or couplings between DE and perfect fluid matter \cite{Boehmer:2015kta}.
Similarly, in the context of Dark Matter (DM), described in the \lcdm model as a pressure-less dust (which can also be thought as a perfect fluid with negligible pressure, although the exact limit of collisions matter strictly speaking does not admit a fluid description), several alternatives have been proposed, mainly motivated by observational challenges that such standard candidate is currently facing at astrophysical scales \cite{Weinberg:2013aya}.
In particular, alternatives to Cold DM such as Warm DM \cite{Colombi:1995ze,Bode:2000gq,Smith:2011ev}, interacting DM \cite{Carlson:1992fn,Spergel:1999mh}, ultralight scalar DM or Bose--Einstein condensate  (BEC) DM \cite{Hu:2000ke,Boehmer:2007um,Suarez:2013iw}, have gained a remarkable attention thanks to their potential ability of curing small scales issues while remaining compatible with the \lcdm on large scales (See \cite{Bertone} for a review of DM candidates). 
Finally, one cannot omit to say that the necessity to introduce such dark components in the Universe energy content seems rooted in the application of GR at scales well beyond those currently test with high precision~\cite{Will:2014xja,Reyes:2010tr}. This has stimulated a novel activity in exploring extended theories of gravitation which often includes extra gravitational degrees of freedom beyond the metric (a prototype of these theories being scalar-tensor theories). 

In this paper we focus on a different approach to extend the standard paradigm. In fact, while all the previous alternatives rely on a modification of the dark component characteristics, or of GR, here we focus on a generalization of the way the former interacts with gravity. In particular, we will consider a perfect fluid, embedded in a curved space-time, that is non-minimally coupled (NMC) to curvature. In doing so we shall show that this turns out not only in a modification of the fluid dynamics but (as expected) in a subtle modification of the gravitational one. It is perhaps important to stress that such a modification of GR dynamics will be associated to a ``coarse grained" regime (the introduction of cosmological perfect fluids and their effective coupling to gravity) rather than to some fundamental modification of gravitational dynamics. As such, it does not rest on the introduction of fundamental degrees of freedom. In this sense this framework is quite different from proposed modifications of the gravitational dynamics aimed at completely avoiding the introduction of dark components. 


The use of fluids in the GR is probably as old as Einstein's theory itself. Their simple and intuitive formalization has made such description of the gravitating matter a fundamental tool in the understanding of the gravitational dynamics and its foundations. Moreover, the theory of fluids has beneficed from its applicability to many different scales and systems \cite{Andersson:2006nr} thus becoming a very well developed framework. However, even in flat space-times, a definitive formulation of a fully relativistic theory of fluids beyond the simplest assumptions is still missing. 
When one moves to the context of cosmology, and hence of GR, the game becomes even more complex as a consequence of the interplay with the gravitational dynamics. 
Indeed, it is still an open question how to extend the Lagrangian formulation of flat space-time theories when moving to curved ones.

Non minimal coupled fluids has been investigated in the attempt of reconciling MOND with DM \cite{Bruneton:2008fk}, in the context of extensions to the dark matter paradigm in \cite{Bettoni:2011fs} and their application to cosmology \cite{Bettoni:2012xv}. Also the above mentioned proposals of DM condensates stimulated the studies of non-minimally coupled relativistic BEC \cite{Bettoni:2013zma}.\footnote{The idea of NMC matter has been also used as an alternative to DM itself \cite{Bertolami:2008im,Bertolami:2008ab}.}

But why a cosmological perfect fluid should be non-minimally coupled to gravity? At the foundation of the fluid paradigm in GR, there is the assumption that the continuous limit (i.e.~the limit from particle to fluid) is reached on scales small enough to be well approximated by flat space time. However, one may wonder what happens if the scale over which the fluid can be defined (i.e. the distance that a particle has to travel in order to have enough collisions to be thermal) is of the same size of the characteristic length over which curvature changes. This is very likely the case for DM, which, being almost non-interacting, has a very large free streaming length so that its mean free path can be of the order of the Hubble scale. More precisely we can estimate it from the formula
\begin{equation}
l_{\rm mfp}=\frac{M_{\textrm{\tiny{DM}}}}{\sigma_{\textrm{\tiny{DM-DM}}}\,\bar{\rho}_{\textrm{\tiny{DM}}}}\sim 10^3 \,\textrm{Gpc}\,,
\end{equation}
where we have taken $\sigma_{\textrm{\tiny{DM-DM}}}/M_{\textrm{\tiny{DM}}}\sim 7\, \textrm{cm}^2/\textrm{g}$ \cite{Dawson:2011kf}, while $\bar{\rho}_{\textrm{\tiny{DM}}}$ is the background DM density. 
With such large mean free path it seems almost natural to include NMC terms in the action. 

A similar reasoning applies to the models based on fluids which have an intrinsic length scale, like the healing length of the BEC DM candidates. Indeed, these models require some form of self-interaction for the dark matter (keeping interactions with standard models field absent or negligible). This might lead to astrophysical scale condensates with potentially very large coherence (healing) lengths which would make dimensionally viable non-minimal coupling terms of the Horndeski type (see \cite{Bettoni:2013zma} for a more detailed discussion).

Besides these heuristic arguments and their application to the DM problems, the idea of a non-minimally coupled fluid is interesting {\it per s\'e} as it provides further insights on the dynamics of fluids in GR. 
In fact, this kind of models show several intriguing and distinctive feature that make the study of their phenomenology interesting. As we shall see a NMC fluid sources the gravitational potential also via the gradients of its density distribution. Moreover, the equations of motion for the fluid show interesting features. In fact, the NMC introduces an extra force in the Euler equation that is related to the gradients of the curvature, while the continuity equation is modified in such a way that it is still possible to define conserved quantities along the flow lines of the fluid.
All these modifications have clear relations with observables and, especially in cosmology, can be easily compared with data.

%

The paper is organized as follows: in section \ref{sec:PFA} we review the theory for a perfect fluid in curved space times while in section \ref{sec:NMCFluid} we extend these results to the case in which the fluid is non-minimally coupled to curvature. In particular we derive the Einstein equations and the fluid equations for both a conformal and a disformal coupling and discuss their properties. In section \ref{sec:cosmo} we analyse the cosmological consequences of the model at both background and linear perturbation level while in section \ref{sec:Newton} we derive the Newtonian limit of the NMC fluid. In section \ref{sec:DE_DM} we discuss the possibility to have more than one fluid which is NMC and the effects that such coupling can have on the DM-DE interaction. Finally, in section \ref{sec:Conclusions} we draw our conclusions.

\section{The action for the relativistic perfect fluid}
\label{sec:PFA}

A fluid is defined to be perfect when it has isotropic pressure, no viscosity, and (in relativity) no heat conduction in the comoving reference frame. As such it can be described by a set of five independent variables, namely the three components of the fluid four velocity $u^\mu$ and by any two scalar functions describing the thermodynamic state of the fluid \cite{landau1959fm,Andersson:2006nr}. However, it is not possible to construct a valid variational principle only with the previous functions unless this variation is constrained \cite{Schutz:1977df}. One possible way to constrain the variation is to introduce Lagrangian multipliers in such a way that the correct fluid equations can be recovered once the properties imposed by those constraints are applied \cite{Brown:1992kc}. 
Our action will be then constructed from a scalar $F$, function of particle number density $n$ and entropy density per particle $s$, i.e. by an equation of state, and from a set of Lagrangian multipliers. The choice of the latter is based on a minimal constrain approach: The variation is constrained in such a way to give the correct equations of motion for a perfect fluid. Practically speaking, this amounts to require the particle number conservation and the absence of entropy exchanges between flow lines.
It will turn out that a more effective description of the action can be done in terms of a vector density field $J^\mu$ defined as
\begin{equation}
J^\mu = \sqrt{-g} \,n u^\mu \,,
\label{eq:Jmu}
\end{equation}
which in particular gives $n=|J|/\sqrt{-g}$ and $u^\mu = J^\mu/|J|$, being respectively a scalar and a pure vector and with $|J|=\sqrt{-g_{\mu\nu}J^\mu J^\nu}$.
The action for the perfect fluid is then \cite{Brown:1992kc}
\begin{equation}
S_{\rm fluid}=\int d^4x\sqrt{-g} F(n,s) + J^\mu\left(\nabla_\mu\varphi+s\nabla_\mu\theta+\beta_A \nabla_\mu\alpha^A\right)\,.
\label{eq:fluidaction}
\end{equation}
The first term is the equation of state of the fluid while the second one it contains the required constraints. The Lagrangian multipliers, $\varphi$, $\theta$ plays the role of thermodynamic potentials and are associated with the chemical free energy and the temperature respectively. The last constraint is instead required to ensure that the flow lines do not cross, i.e. there is a one-to-one mapping between the Lagrangian and the Eulerian coordinates at any two fixed spatial hyper-surfaces. In particular, the $\alpha^A$ and $\beta_A$, $A=1,2,3$, are the Lagrangian coordinates of the fluid.
This two part structure has the particular advantage that it divides the action into a metric dependent and metric independent part. This help in distinguishing the properties of the embedding of the fluid in a curved background from the symmetries that must be enforced independently of the geometry.

In fact, the variation with respect to the metric gives a stress energy tensor (SET) for the fluid
\begin{equation}
T^{\mu\nu}_{\mathrm{fluid}}\equiv \frac{2}{\sqrt{-g}}\frac{\delta S_{\mathrm{fluid}}}{\delta g_{\mu\nu}}=Fg^{\mu\nu}-n\frac{\partial F}{\partial n}(u^\mu u^\nu+g^{\mu\nu})\,.
\label{SET_F}
\end{equation}
This has the form of a perfect fluid SET and defining $\rho\equiv T^{\mu\nu}u_\mu u_\nu$ and $p\equiv T^{\mu\nu} h_{\mu\nu}/3$ it can be cast in the standard form $T^{\mu\nu}=(\rho+p)u^\mu u^\nu+pg^{\mu\nu}$ with the identifications
\begin{equation}
\rho\equiv -F\,, \qquad p=n \frac{\partial F}{\partial n} -F\,.
\label{eq:PF_F_relation}
\end{equation}
This also tell us that the Lagrangian for the perfect fluid is the energy density measured by an observer comoving with the fluid.\footnote{Equivalent Lagrangian densities can be obtained adding surface terms and using the thermodynamic relations for the fluid.}
Adding the standard GR kinetic term for the metric and varying it with respect to the metric gives the Einstein equation
\begin{equation}
\mpl^2 G^{\mu\nu}= T_{\mathrm{fluid}}^{\mu\nu}\,.
\end{equation}
The equation of motion for the perfect fluid can be obtained by taking the covariant derivative of the Einstein equations. With this procedure we find the standard continuity and Euler equations
\begin{eqnarray}
&& \dot{\rho}+\theta(\rho+p)=0\,,\\
&&\left(\rho + p \right) \dot{u}^\sigma +h^\sigma{}_\nu\nabla^\nu p=0\,.\label{eq:Euler_MC}
\end{eqnarray}
where we have defined $\dot{()}=u^\mu\nabla_\mu()$ and $\theta=\nabla_\mu u^\mu$.
It is interesting to note that this is a set of four equations for five variables. Usually, this issue is solved by assuming an equation of state for the fluid which relates the pressure to the density. However, recalling that $\rho=\rho(n,s)$, the continuity equation can be expanded as
\begin{equation}
\frac{\partial\rho}{\partial n}(\dot n+\theta n)+\frac{\partial\rho}{\partial s}\dot s=0\,.
\end{equation}
We now see that there is one missing equation to close the system. This is provided by the request that the particle number density is conserved along the flow. This amounts to say that the first parenthesis is independently zero so that the conservation of entropy follows automatically and the system is closed \cite{Misner:1974qy}.

Until now, no use of the constraints has been done. However, we will discuss now how their role is fundamental in order to obtain the correct equations of motion for  the fluid when those are derived by varying the action with respect to the fluid variables. In fact, such variation leads to the following set of equations
\begin{eqnarray}
0 &=& \frac{\delta S}{\delta J^\mu}=\mu U_\mu +\nabla_\mu\varphi +s\nabla_\mu\theta +\beta_A\nabla_\mu\alpha^A\,,\label{eq:EulerPF1}\\
0 &=& \frac{\delta S}{\delta \varphi}=-\nabla_\mu J^\mu\,,\label{eq:continuityPF1}\\
0 &=& \frac{\delta S}{\delta \theta}=-\nabla_\mu(sJ^\mu)\,,\label{eq:entropyPF1}\\
0 &=& \frac{\delta S}{\delta s}=\sqrt{-g}\frac{\partial F}{\partial s}+J^\mu\nabla_\mu\theta\,,\\
0 &=& \frac{\delta S}{\delta \alpha^A}=-\nabla_\mu(\beta_AJ^\mu)\,,\\
0 &=& \frac{\delta S}{\delta \beta_A}=J^\mu\nabla_\mu\alpha^A\,.
\label{eq:constraints}
\end{eqnarray}
According to the definition of $J^\mu$ the continuity equation is obtained by combining together the equation for the entropy \eqref{eq:entropyPF1} and number density conservation \eqref{eq:continuityPF1} while the Euler equation is obtained by differentiating equation \eqref{eq:EulerPF1} with respect to $\nabla_\nu$, then anti-symmetrizing it and projecting it along the mixed direction $u^\mu h^{\sigma\nu}$ \cite{Brown:1992kc}. 

The fluid equations can be rephrased in a more thermodynamic form contracting the first equation with the four velocity $u^\mu$ and using the relation between $J^\mu$ and the other fluid variables \eqref{eq:Jmu}
\begin{eqnarray}
\mu &=& f+Ts\,,\\
\nabla_\mu(nu^\mu)&=&0\,,\\
u^\mu\nabla_\mu s &=&0\,,\\
\frac{1}{n}\frac{\partial \rho}{\partial s}&=&T\,,\\
u^\mu\nabla_\mu\beta_A&=&0\,,\\
u^\mu\nabla_\mu\alpha^A&=&0\,,
\end{eqnarray}
where $\mu = (p+\rho)/n$, $f= (p+\rho)/n -Ts$. In this form the equations give the relation between enthalpy $\mu$, temperature $T$ and free chemical energy $f$ and express the conservation of number density and of entropy while the last two equation are more of geometrical nature and tells us that the Lagrangian coordinates are constant along the flow lines.

As a final remark we discuss in more detail the role of the two parts of the fluid Lagrangian. The fluid being perfect is related to the fact that in its metric dependent part no derivatives of the fluid variables appear. Hence, its SET can only be of the perfect fluid form. The role of the constraints is to enforce the properties of the fluid we know from the conservation of the SET when doing the variation with respect to the fluid variables. Of course, the two parts are related in order to give a consistent theory so, for example, inclusion of diffusion terms will need a change in both parts of the Lagrangian. Hence, it is not trivial how to generalize this action to more complex situations. We will see in the next paragraphs how one can obtain a more general action for a fluid still compatible with the constraints.

\section{Non-minimally coupled fluid}
\label{sec:NMCFluid}

As we have discussed in the introduction, the minimally coupled perfect fluid paradigm may not be well motivated in GR for fluids with mean free paths of the same order of the curvature scale and we have pointed out how an effective coupling between fluid variables and curvature may rise. 

However, when moving from flat to curved space-times, there is no general prescription on the kind of couplings between curvature and fluid variables and their derivatives so that an a priori selection criterion must be adopted. Here we take advantage of the natural derivative structure of fluids to argue that there are only two non-minimal couplings that can be added to the perfect fluid Lagrangian without introducing derivatives of the fluid variables:\footnote{Modulo surface terms. Notice also that by derivatives of the fluid we mean terms that cannot be eliminated via integration by parts.
}  a conformal coupling, $F_c(n,s)R$, and a disformal one, $F_d(n,s)R_{\mu\nu}u^\mu u^\nu$.\footnote{One may argue that, as curvature terms contain derivatives of the metric, such NMC terms may be thought as fluid derivatives contributions hence requiring an extension of the model to include all possible derivative terms up to second order. However, those terms manifestly break the perfect fluid assumption and will for this reason not included.}

Hence, the most general action that contains non-minimal couplings and is still zero order in the fluid derivatives is \cite{Bettoni:2011fs}
\begin{equation}
S=\frac{\mpl^2}{2}\int d^4 x\sqrt{-g}\big[R+\as F_c(n,s)R+\ar F_d(n,s) R_{\mu\nu}u^\mu u^\nu\big]+S_{\mathrm{fluid}}\,,
\end{equation}
where $S_{\mathrm{fluid}}$ is given by \eqref{eq:fluidaction}.
Notice that the associated total Lagrangian can be rewritten as
\begin{equation}
\mathcal{L}=\frac{\mpl^2}{2}R+F(n,s)\left[1+\frac{\mpl^2}{2}\left(\as\frac{F_c(n,s)}{F(n,s)} R+\ar \frac{F_d(n,s)}{F(n,s)}R_{\mu\nu}u^\mu u^\nu\right)\right]\,,
\end{equation}
which can be seen as a shift in the equation of state of the minimally coupled fluid.  In the simplest case of $F_c=F$ and $\ar=0$ the shift is constant from the fluid variable point of view and hence we expect to see changes in the equation of motion for the fluid to appear as shifts and rescaling (See appendix \ref{app:FEOM}) . Of course, the Einstein equations will be modified more significantly. In this sense, there will be differences in the conserved fluid quantities and the ones sourcing the gravitational potentials. However, as we shall discuss in more details below which one to choose will be mostly a matter of convenience. 

Before moving the the details we note that the presence of a non-minimal coupling in the action will produce a coupling between second derivatives of the metric and the fluid in the Einstein equations so that there is no unique way to define, from the metric equations, the SET components as seen from an observer. In fact, the Einstein equations will take the general form
\begin{equation}
M_\textrm{Pl}^2(\psi)G_{\mu\nu}=T_{\mu\nu}^{\textrm{eff}}(R,\psi,\partial\psi,\partial\partial \psi)\,,
\end{equation}
where $\psi$ stands for fluid variables and where $T_{\mu\nu}^{\textrm{eff}}$ contains both the minimally coupled perfect fluid SET and the corrections due to the NMC. One may then envisage at least three different schemes for writing these equations:
\begin{itemize}
\item Define an effective Planck mass $M_*^2$ and an effective SET;
\item Fix the Planck mass to be constant and define an effective SET that will contain the Einstein tensor;
\item Use contractions of the Einstein equations with the metric and the fluid four velocity to get rid of the curvature on the right hand side. However, the presence of second order derivatives of fluid variables will still mix gravitational and fluid degrees of freedom.
\end{itemize}

Of course there is no physics into this and the choice of which prescription to use will be mostly determined by the observables chosen to compare the model with data or by the numerical procedure used to integrate the equations.\footnote{A physical definition of the effective Planck mass can be given by looking at the coefficient of the graviton kinetic term appearing in front of the second derivative term in the expansion of the action for perturbations around some background.} Here, we will pick the first possibility which was recently used in \cite{Bellini:2014fua} so that the Einstein equations are
\begin{equation}
M_*^2 G_{\mu\nu}= T^{\mathrm{eff}}_{\mu\nu}\,,
\end{equation}
while the fluid equations are derived by taking the covariant derivative of such equation, 
\begin{equation}
\nabla_\nu T_{\mathrm{eff}}^{\mu\nu} = G^{\mu\nu}\nabla_\nu M_*^2\,.
\end{equation}
Notice that, being in the Jordan frame, any other matter species which is minimally coupled will be separately conserved \cite{Koivisto:2005yk}.

Finally, as a consistency check, in appendix \ref{app:FEOM} we have also computed the fluid equations by varying the action with respect to the fluid variables, including the perfect fluid constraints, and found the same equations obtained from the conservation of the Einstein equations.

\subsection{Conformally coupled perfect fluids}
\label{sec:CCF}

As a first investigation we consider the case of a perfect fluid which is coupled to the Ricci scalar. In this case the action is
\begin{equation}
S_C=\frac{\mpl^2}{2}\int d^4x\sqrt{-g}\big[1+\as F_c(n,s)\big]R+S_{\rm fluid}\,,
\label{eq:CCFA}
\end{equation}
where $S_{\rm fluid}$ is given in \eqref{eq:fluidaction} and $F_c(n,s)$ is a new arbitrary function of the fundamental fluid variables.

The variation with respect to the metric gives the Einstein equations for the conformally coupled fluid
\begin{equation}\label{eq:EFEconf}
(\mpl^2 +\as\mpl^2 F_c)G_{\mu\nu}-g_{\mu\nu}F+h_{\mu\nu}n\left(\frac{\partial F}{\partial n}+\as\frac{\mpl^2}{2} R\frac{\partial F_c}{\partial n}\right)+\as\mpl^2\left(g_{\mu\nu}\Box F_c-\nabla_\mu\nabla_\nu F_c\right)=0\,.
\end{equation}
The contributions coming from the NMC appear as a modification to the Planck mass that is now fluid dependent, as a correction to the perfect fluid structure of the SET and as a derivative contribution.
At this point one may want to follow the same strategy used for the minimally coupled case in order to identify the thermodynamic quantities of the fluid, like energy density and pressure.
However, 
 now there is no general prescription that allows one to unequivocally define the fluid variables from the Einstein equations. According to the discussion presented above, we then define an effective Planck mass 
\begin{equation}
M_*^2\equiv \mpl^2(1 +\as F_c)\,,
\label{EPMC}
\end{equation}
 and an effective SET
\begin{equation}
T_{\mu\nu}^\eff=g_{\mu\nu}F-h_{\mu\nu}n\left(\frac{\partial F}{\partial n}+\as\frac{\mpl^2}{2} R\frac{\partial F_c}{\partial n}\right)-\as\mpl^2\left(g_{\mu\nu}\Box F_c-\nabla_\mu\nabla_\nu F_c\right)\,,
\label{EFEC}
\end{equation}

 so that the Einstein equations are
\begin{equation}
M_*^2 G_{\mu\nu}= T^{\mathrm{eff}}_{\mu\nu}\,,
\end{equation}
where $M_*^2$ and $T^{\mathrm{eff}}_{\mu\nu}$ are given in \eqref{EPMC} and \eqref{EFEC} respectively.

The fluid equations can be derived by taking the covariant derivative of the Einstein field equations. After some algebra we get
\begin{equation}
\nabla_\nu F+\as\frac{\mpl^2}{2} R\nabla_\nu F_c-n\left(\frac{\partial F}{\partial n}+\as\frac{\mpl^2}{2} R\frac{\partial F_c}{\partial n}\right)(\dot{u}_\nu+\theta u_\nu)+h^\mu{}_\nu\nabla_\mu\left(n\frac{\partial F}{\partial n}+\as\frac{\mpl^2}{2} Rn\frac{\partial F_c}{\partial n}\right)=0\,.
\end{equation}
Projecting this equation along $u^\nu$ gives
\begin{equation}
\left(\frac{\partial F}{\partial n}+\as\frac{\mpl^2}{2} R\frac{\partial F_c}{\partial n}\right)(\dot n+\theta n)+\left(\frac{\partial F}{\partial s}+\as\frac{\mpl^2}{2} R\frac{\partial F_c}{\partial s}\right)\dot s=0\,.
\label{eq:continuity_C}
\end{equation}
As for the case of the minimally coupled  perfect fluid, we need to impose an extra equation in order to close the system. The structure of equation \eqref{eq:continuity_C} and its similarity to the analogous one for the case of a minimally coupled fluid suggest that we take again as extra equation the conservation of the particle number density along the flow. Such imposition is also the one that makes compatible the above system of equations with the one derived in appendix \ref{app:FEOM}.\footnote{This is not a completely arbitrary choice. Indeed, the coupling between the variables $n$ and $s$ and curvature is `scalar' in the sense that these variables are coupled to scalar quantities, namely $R$ and $\langle R\rangle$ and hence we do not expect in the equation of motion terms able to generate a spatial particle flow. The situation would be different if we had considered a NMC of the form $R_{\mu\nu}\nabla^\mu F(n,s) \nabla^\nu F(n,s)$ which has a clear direction dependence.} As a consequence of this, any scalar function of $n$ and $s$ will obey the following equation
\begin{equation}
\dot{\mathcal{F}}+\theta n\frac{\partial \mathcal{F}}{\partial n}=0\,,
\end{equation}
with $\mathcal{F}=\mathcal{F}(n,s)$. In analogy with the minimally coupled perfect fluid this allows us to write a formal continuity equation defining a density $\rho_\mathcal{F}\equiv -\mathcal{F}$ and a pressure $p_\mathcal{F}\equiv -n\partial \mathcal{F}/\partial n +\mathcal{F}$ so that
\begin{equation}
\dot{\rho}_\mathcal{F}+\theta (\rho_\mathcal{F}+p_\mathcal{F})=0\,.
\label{eq:general_continuity}
\end{equation}
This is interesting as it implies that even with the NMC it is possible to identify quantities that are conserved along the flow line.\footnote{Interestingly, an analogous result has been found in the case of perfect fluid coupled to a scalar field \cite{Boehmer:2015kta}.} In particular, $F(n,s)$ and $F_c(n,s)$ both (or any combination of them) are conserved and hence can be used to define fluid variables.

The projection onto the orthogonal hypersurface gives
\begin{multline}
n\left(\frac{\partial F}{\partial n}+\as\frac{\mpl^2}{2} R\frac{\partial F_c}{\partial n}\right)\dot u^\sigma-\\
h^{\sigma\nu}\nabla_\nu\left(F-n\frac{\partial F}{\partial n}-\as\frac{\mpl^2}{2}\left(\frac{\partial F_c}{\partial n}n-F_c\right)R\right)+\as\frac{\mpl^2}{2} F_c h^{\sigma\nu}\nabla_\nu R=0\,.
\label{eq:Euler_CCF}
\end{multline}
We can see that the NMC has the effect of modifying the weight function suppressing the spatial gradients sourcing the time derivative of the fluid velocity, the pressure contribution and add a new potential related to the local value of the Ricci scalar. Schematically we can rewrite the previous equations in a form that is closer to the one for a minimally coupled fluid \eqref{eq:Euler_MC}, namely
\begin{equation}
\left(\rho_{\textrm{tot}}+p_{\textrm{tot}}\right)\dot u^\sigma-h^{\sigma\nu}\nabla_\nu\left(p_{\textrm{tot}}\right)-\as\frac{\mpl^2}{2} \rho_c h^{\sigma\nu}\nabla_\nu R=0\,,
\end{equation}
where $\rho_c\equiv -F_c$ while $p_{\textrm{tot}}$ and $\rho_{\textrm{tot}}$ are collective labels for the various coefficients appearing  in equation \eqref{eq:Euler_CCF} and read
\begin{eqnarray}
\rho_{tot}&=&F+\frac{\as}{2}\mpl^2R F_c\,,\\
p_{tot}&=&n\frac{\partial \rho_{tot}}{\partial n}-\rho_{tot}\,.
\end{eqnarray}
This is interesting as it shows that even in the simple case of conformally coupled fluids, the NMC has a clear and distinguishable feature: It introduces an extra force term. In fact, while the pressure renormalization can be in principle mimicked with other exotic fluid, the presence of a new force term related to the local value of the Ricci scalar, is a signature of this model. 
By means of the trace of the Einstein equations and taking terms up to $\mathcal{O}(\as)$, we can relate this term to the gradients of trace of the SET of all the matter fields. Hence, this force will be effective not only in the presence of gradients of the NMC fluid but whenever any matter distribution is sufficiently inhomogeneous.

%

\subsection{Disformally coupled perfect fluids}
\label{sec:DCF}
We now consider the case in which the NMC is of the form $F_d(n,s)R_{\mu\nu}u^\mu u^\nu$. The action for this model is
\begin{equation}
S_D=\int d^4x\sqrt{-g}\left[\frac{\mpl}{2}\left(R+\ar F_d(n,s)\enangle{R}\right)\right]+S_{\rm fluid}\,,
\label{eq:DCFA}
\end{equation}
where $S_{\rm fluid}$ is given in \eqref{eq:fluidaction} and $F_d(n,s)$ is a new arbitrary function of the fundamental fluid variables, while $\enangle{R}\equiv R_{\mu\nu}u^\mu u^\nu$.
Even in this case no derivatives of the fluid variables are present in the action. Here, however, one could re-express the NMC coupling using the commutation rule of the covariant derivative 
\begin{equation}
R_{\mu\nu}u^\mu u^\nu=u^\mu\left(\nabla_\nu\nabla_\mu u^\nu-\nabla_\mu\nabla_\nu u^\nu\right)\,,
\end{equation}
so that an apparent higher order derivative term has appeared. However, given that this combination is zero in flat space time it is more reasonable to consider such interaction as a gravity-fluid coupling rather than a thermodynamic property of the fluid. 

Taking the metric variation one gets the following Einstein equations for the disformal fluid
\begin{multline}
 \mpl^2G_{\mu\nu} - F g_{\mu\nu} + \bigl(g_{\mu\nu} + u_{\mu} u_{\nu}\bigr) n \frac{\partial F}{\partial n}+ \ar\frac{\mpl^2}{2} \Bigl(- {F_d} \enangle{R} (g_{\mu\nu} + 2 u_{\mu} u_{\nu})-\\
 \left.2 \nabla_\alpha\nabla_{(\mu} t_{\nu)}{}^{\alpha} + \Box t_{\mu\nu} + g_{\mu\nu} \nabla_\alpha\nabla_\beta t^{\alpha\beta} + g_{\mu\nu} \enangle{R} n \frac{\partial F_d}{\partial n} + \enangle{R} u_{\mu} u_{\nu} n \frac{\partial F_d}{\partial n}\right)=0\,,
\label{eq:EFEdisf}
\end{multline}
where the round brackets stand for symmetrized of the indices and where we have introduces for simplicity the tensor
\begin{equation}
t^{\mu\nu}=F_du^\mu u^\nu\,.
\end{equation}
Notice that in this case the Planck mass is constant so that in principle we could define  the observed fluid variables by decomposing the Einstein equations in its effective variables but again this would hide in the matter variables curvature terms that will ultimately mix gravitational and fluid degrees of freedom. For this reason and to be consistent with the formulation of the conformally coupled case we stick with our notation. We then have a constant Planck mass $ M_*^2=M_{\mathrm{Pl}}$ and an effective SET
\begin{eqnarray}
\nonumber
T^{\mathrm{eff}}_{\mu\nu} &=&  F g_{\mu\nu} - \bigl(g_{\mu\nu} + u_{\mu} u_{\nu}\bigr) n \frac{\partial F}{\partial n}- \ar\frac{\mpl^2}{2} \Bigl(- F_d \enangle{R} (g_{\mu\nu} + 2 u_{\mu} u_{\nu}) \\
&+& \left. \nabla_\alpha\nabla_{(\mu} t_{\nu)}{}^{\alpha} - \Box t_{\mu\nu} + g_{\mu\nu} \nabla_\alpha\nabla_\beta t^{\alpha\beta} - g_{\mu\nu} \enangle{R} n \frac{\partial F_d(n)}{\partial n} - \enangle{R} u_{\mu} u_{\nu} n \frac{\partial F_d}{\partial n}\right)\,,
\end{eqnarray}
so that the Einstein equation for the disformally couple fluid are
\begin{equation}
M_{\mathrm{Pl}}^2 G_{\mu\nu}= T^{\mathrm{eff}}_{\mu\nu}\,.
\end{equation}
where the effective SET is given by the expression above. 
We now move to the computation of the fluid equations of motion. Again we take the covariant derivative of the Einstein equations and project the result along the direction parallel to the fluid four velocity and onto its orthogonal hypersurface. 
Quite remarkably for the first one we get
\begin{equation}\label{eq:Euler_DCF}
\left(\frac{\partial F}{\partial n} + \ar \frac{\mpl^2}{2} \frac{\partial F_d}{\partial n}\enangle{R}\right)\left(\dot{n}+\theta n\right)+\left(\frac{\partial F}{\partial s} + \ar\frac{\mpl^2}{2} \frac{\partial F_d}{\partial s}\enangle{R}\right)\dot{s}=0\,,
\end{equation}
which again tells us that the particle number density and entropy per particle are conserved along the flow lines, with the conclusion that any scalar function of these variable obeys the continuity equation \eqref{eq:general_continuity}. Even if surprising, this result is in line with the expectation that our NMC action is compatible with the perfect fluid assumptions.
The spatial projection of the fluid equations gives the modified Euler equation
\begin{multline}
\left[n\frac{\partial F}{\partial n} -\ar \enangle{R}\left(2 F_d-n \frac{\partial F_d}{\partial n}\right)\right]\dot u^\sigma+ h^{\sigma\gamma}\nabla_\gamma\left[n\frac{\partial F}{\partial n}-F +\ar\left(n\frac{\partial F_d}{\partial n}-F_d\right)\enangle{R}\right]\\ 
+F_du^\alpha u^\beta h^{\sigma\gamma}\nabla_\gamma R_{\alpha\beta} =\ar\left[2\theta\left(F_d-n\frac{\partial F_d}{\partial n}\right)h^{\sigma\gamma}R_{\gamma\beta}u^\beta+2F_d h^{\sigma\gamma}u^\alpha\nabla_\alpha\left(R_{\gamma\beta}u^\beta\right)\right]\,.
\end{multline}
This equation is more complicated than that for a conformally coupled fluids, but still we can see some common features. There is a modified coefficient in front of the time derivative of the fluid four velocity, a modified pressure term and an extra force related to the Ricci tensor. The new feature of this case is that on the right hand side of the equation we have two terms that act as source for the equation. These terms represent a new kind of contribution which is not related to spatial gradients and hence represents another novel feature of the NMC. Notice also that any component of the acceleration $\dot u^\sigma$ will in general receive contributions from all the others via the contraction $h^{\sigma\beta}R_{\beta_\gamma}\dot u^\gamma$ present in the last term of the second line.

Finally, we note that the last term in the second line, despite the appearance of a derivative of the Ricci tensor along the flow line of matter, is not introducing any new degree of freedom. In fact, it can be rewritten as
\begin{equation}
h^{\sigma\gamma}u^\alpha \nabla_\alpha\left(R_{\gamma\beta}u^\beta\right)=h^{\sigma\gamma}u^\alpha \left(\nabla_\alpha\nabla_\gamma\nabla_\beta u^\beta-\nabla_\beta\nabla_\gamma u^\beta\right)\,,
\end{equation}
which shows that of the three derivatives at most two are time derivatives so that only second time derivatives of the metric will appear.

%
%

\subsection{Count of degrees of freedom}
\label{sec:dof}

In the previous sections we have seen how the coupling of fluid variables to curvature terms implies the appearance of second derivatives of such variables in the Einstein equations. Hence, an important point to address in this class of models is whether the NMC introduces unwanted degrees of freedom. Stated in another way, one needs to check that
the standard relativistic hydrodynamics is not spoiled when introducing such
non-minimal couplings. 

It is the aim of this section to discuss this point for the two cases of conformally and disformally coupled perfect fluid, providing arguments that point towards the absence of such extra degrees of freedom.

\subsubsection{Conformally coupled perfect fluid}

In the case of conformally coupled perfect fluids the answer is straightforward. The equations for this case are \eqref{eq:EFEconf}, \eqref{eq:Euler_CCF}, \eqref{eq:constraints_CCF_number} and \eqref{eq:constraints_CCF_entropy}. In the Einstein equations, second derivatives of the function $F_c$ appear which may lead to unwanted second time derivatives of the fluid variables $n$ and $s$. However, the conservation of the particle number density and of entropy, the last two equations of the system, implies that we can always eliminate such derivatives and hence no extra degrees of freedom are introduced in this model.

\subsubsection{Disformally coupled perfect fluid}

In the case of disformally coupled perfect fluids the situation is more complex. In fact, in the Einstein equations \eqref{eq:EFEdisf} appear derivative of the tensor $t_{\mu\nu}=F_d(n,s)u_\mu u_\nu$. When those are expanded they will give second derivatives of the scalar functions $n$ and $s$ as well as of the fluid four velocity. The former can be eliminated by the same arguments used for the case of a conformally coupled fluids as also in this case conservation equations holds for those variables as can be seen from equations \eqref{eq:constraints_DCF_number} and \eqref{eq:constraints_DCF_entropy}. However, such constraint does not exist for the fluid four velocity. It is hence important to check that the Einstein equations are actually free of second time derivatives of the fluid velocity.
In order to show this we will make use of the natural space-time foliation provided by the vector field and project the Einstein equations accordingly.\footnote{This is the standard procedure to define the observed energy density, momentum and stresses of a system as seen by an observer whose world-line is defined by a vector $u^\mu$.}
\paragraph{Time projection}
Projecting the Einstein equations along the tensor $u^\mu u^\nu$ it gives the following scalar equation
\begin{equation}
\mpl^2\langle G\rangle +F -\frac{\ar}{2}\mpl^2 F_d \langle R\rangle+\frac{\ar}{2}\mpl^2\left[-2u^\mu u^\nu\nabla_\alpha\nabla_\mu t_\nu{}^\alpha+ u^\mu u^\nu\Box t_{\mu\nu}-\nabla_\alpha \nabla_\beta t^{\alpha\beta}\right]=0
\end{equation}
We now proceed with the analysis of the last parenthesis which is the one containing the second derivatives of the fluid variables. As a consequence of the unit norm of the four velocity, the following relations holds
\begin{equation}\label{eq:unit_constraints}
u^\mu\nabla_\alpha u_\mu=0\,,\qquad u^\mu\nabla_\alpha\nabla_\beta u_\mu=-\nabla_\alpha u^\mu \nabla_\beta u_\mu\,,
\end{equation}
and hence, by expanding the derivatives in the last parenthesis and keeping only second derivatives of the velocity, we get
\begin{equation}
-2u^\mu u^\nu\nabla_\alpha\nabla_\mu t_\nu{}^\alpha+ u^\mu u^\nu\Box t_{\mu\nu}-\nabla_\alpha \nabla_\beta t^{\alpha\beta}\sim 2\left(\dot u^\nu\right)^2+2\left(\nabla_\alpha u_\beta\right)^2 + \langle R\rangle\,.
\end{equation}
Hence, no second derivatives of the four velocity are present.

\paragraph{Space-time projection}

We now project the Einstein equations along the mixed direction $u^\mu h^{\sigma\nu}$, where $h$ is the spatial projector onto the hypersurface orthogonal to the fluid velocity. This gives
\begin{equation}
h^{\sigma\nu}u^\mu G_{\mu\nu} +\frac{\ar}{2}\left[-2h^{\sigma\nu}u^\mu \nabla_\alpha\nabla_{(\mu}t_{\nu)}{}^\alpha + h^{\sigma\nu}u^\mu\Box t_{\mu\nu}\right]=0\,.
\end{equation}
Again, considering only the derivative term and keeping only the second derivatives we get
\begin{equation}
-2h^{\sigma\nu}u^\mu \nabla_\alpha\nabla_{(\mu}t_{\nu)}{}^\alpha + h^{\sigma\nu}u^\mu\Box t_{\mu\nu} \sim h^{\sigma\nu}\left[-h^{\alpha\beta}\nabla_\alpha\nabla_\beta u_\nu + \dot u^\mu\nabla_\nu u_\mu +\nabla_\nu \theta\right]\,,
\end{equation}
which, again, does not contain any second time derivative of the four velocity.

\paragraph{Spatial projection}

This time we project the Einstein equation onto the orthogonal subspace. This amount to contract the Einstein equations with the tensor $h^{\sigma\mu}h^{\rho\nu}$. The result is
\begin{multline}
\mpl^2 h^{\sigma\mu}h^{\rho\nu} G_{\mu\nu}+\left(-F+n\frac{\partial F}{\partial n}\right) h^{\sigma\rho}+\frac{\ar}{2}\mpl\langle R\rangle\left[-F_d+\frac{\partial F_d}{\partial n}\right]h^{\sigma\rho}\\
+\frac{\ar}{2}\mpl^2 h^{\sigma\mu}h^{\rho\nu}\left[-2\nabla_\alpha\nabla_{(\mu}t_{\nu)}{}^\alpha+\Box t_{\mu\nu} + g_{\mu\nu}\nabla_\alpha\nabla_\beta t^{\alpha\beta}\right]=0\,.
\end{multline}
Again, extracting only the second derivative from the last parenthesis of the above equation leads to the following terms
\begin{equation}
h^{\sigma\mu}h^{\rho\nu}\left[-2\nabla_\alpha\nabla_{(\mu}t_{\nu)}{}^\alpha+\Box t_{\mu\nu} + g_{\mu\nu}\nabla_\alpha\nabla_\beta t^{\alpha\beta}\right] \sim -h^{\sigma\mu}h^{\rho\nu} u^\alpha\nabla_\alpha\left(\nabla_\mu u_\nu +\nabla_\nu u_\mu\right) +2 h^{\sigma\rho}\dot\theta\,.
\end{equation}
The last term does indeed contain second derivatives of the fluid four velocity. In fact $\dot \theta = u^\alpha \nabla_\alpha \nabla_\beta u^\beta$. However, thanks to the unit norm constraints \eqref{eq:unit_constraints}, it is easy to show that
\begin{equation}
\theta = h^{\alpha}{}_\mu\nabla_\alpha u ^\mu\,,
\end{equation}
which tells us that the expansion $\theta$ reduces to a spatial derivative of the fluid four velocity. Hence, the term $\dot \theta$ actually contains only one time derivative. We thus conclude that, as for the previous components, also in this case no higher time derivatives of the fluid four velocity are generated by the NMC.

In this section we have made a first analysis of the number of degrees of freedom for a NMC fluid and concluded that despite the appearances no extra degrees of freedom seems to be generated by the coupling. Of course, only an Hamiltonian analysis can say the final word on this aspect but, as we will see below, our intuitions are confirmed by the analysis of the linearly perturbed equations around a Friedman--Lama\^itre-Robertson--Walker (FLRW) background. It is indeed very likely that the results derived in this section will hold also when a more complete analysis will be done.

\section{Cosmology of a non-minimally coupled perfect fluid}
\label{sec:cosmo}

A crucial test for all the modified gravity models comes from their ability to match the background evolution of the Universe and the process of structure formation. Even if a detailed analysis of the cosmology is beyond the scopes of this work, we here derive the main equations and discuss the modifications produced by the NMC. To do so we consider a FLRW metric defined by the usual squared line element 
\begin{equation}
ds^2 = -(1+2\Phi)dt^2 + a^2(t)(1-2\Psi) d{\bf x}\cdot d{\bf x}\,,
\end{equation}
 and apply it to the Einstein equations \eqref{eq:EFEconf} and \eqref{eq:EFEdisf} respectively as well as to the fluid equations \eqref{eq:general_continuity}, \eqref{eq:Euler_CCF} and \eqref{eq:Euler_DCF}.

\subsection{Cosmological background}
\label{sec:bckgr}

We start our investigation by looking at the equations for the background, setting the gravitational potentials $\Phi$ and $\Psi$ to zero. As a first step we recall that from \eqref{eq:general_continuity} it can be inferred that the background evolution of the fluid equation of state $F$ as well as that of the NMC functions $F_c$ and $F_d$ is determined once specific equation(s) of state are taken. In particular, we have
\begin{equation}
\dot{\rho}_i=-3H(\rho_i+p_i)\,,
\end{equation}
with $\rho_i=-F_i$ and $p_i=F_i-\partial F_i/\partial n$. Recall that the NMC functions are dimensionless and hence their density and pressure are to be considered as divided by some reference scale. Finally, as it will be used later, we derive also the expression for the second time derivative of the density
\begin{equation}
\ddot{\rho}_i=\left[-3\dot H+9H^2 (1+c_i^2)\right](\rho_i+p_i)\,,
\end{equation}
where $c_i^2 =\partial p_i/\partial\rho_i$ is the speed of sound for the i-th species. For simplicity we further assume that the minimally coupled fluid has negligible pressure. In this limit, equation  \eqref{eq:PF_F_relation} implies $F\propto n$ so that one can expect that $F_c(n,s)=F_c(F,s)$ as well as $F_d(n,s)=F_d(F,s)$. With the identification $F=-\rho$, being the latter the minimally coupled energy density, one can deduce that the NMC function will be generic functions of $\rho$ \textbf{and $s$}. 

With these details at hand we now proceed with the investigation of the cosmological background evolution of our NMC model. 

\subsubsection{Conformally coupled fluid}

For the conformal NMC the Einstein equations reduces to the following system
\begin{equation}
H^2 = \frac{8\pi G}{3}\frac{ \rho}{1+\as(2\rho_c+3p_c)} \,,
\end{equation}
\begin{equation}
 \dot H  =- 4\pi G \rho \frac{1 + \as(p_c-3(\rho_c+p_c)c_c^2)}{\Bigl(1 + \as(2\rho_c+3 p_c)\Bigr)^2}\,.
\end{equation}
From these equations we see that the effects of the NMC reduce to a rescaling of the minimally coupled density $\rho$.\footnote{If we want to include other minimally coupled matter species it is sufficient to replace $\rho$ with the sum of all the other components. This also tells us the universal nature of the NMC as its modifications affects all the matter species.} It is important to notice that the denominator $1+\as(2\rho_c+3p_c)$ has to be strictly positive in order to avoid singularities and the flip the gravitational constant sign. Of particular interest is the sign of the expression between brackets in the second equation. If the NMC terms start to dominate at late times they can flip the sign of $\ddot{a}$, thus producing a positive acceleration even if the minimally coupled matter species satisfy the strong energy condition.

However, in order to reproduce the observed matter and radiation dominated eras, the NMC contributions have to become negligible as we move backwards in time. This can be easily achieved if, for example, the equation of state of the NMC fluid is such that the density $\rho_c$ is a  growing function of the scale factor. 

\subsubsection{Disformally coupled fluid}

For the disformal NMC the Einstein equations are
\begin{equation}
H^2 = \frac{8\pi G}{3} \frac{\rho}{1 - \frac{1}{2} \ar( \rho_d +3p_d)}\,,
\end{equation}
\begin{equation}
\dot H  =-4\pi G\rho \frac{1+\ar(p_d-\frac{3}{2}(\rho_d+p_d)c_d^2)}{(1 - \frac{1}{2} \ar( \rho_d +3p_d))^2} \,.
\end{equation}
Similarly to what we found for the conformal case, the NMC rescales the energy density contribution and, in particular, we have to require that the denominator in the two equations is never zero. 
In the second equation, the denominator is always positive while the numerator can freely change its sign so that also in this case the NMC may change the sign of the acceleration.

\subsection{Linear perturbations}
\label{sec:linpert}

We now proceed with the investigation of the dynamics of linear perturbations. We define the density perturbation as $\rho(t,\vec x)= \bar \rho(t) +\delta\rho(t, \vec x)$. Thanks to the conservation of both number density and entropy, the perturbations in the densities $\rho$, $\rho_c$ and $\rho_d$ obey the following equation
\begin{equation}
\dot{\delta\rho}_i+3H\left(1+c_i^2\right)\delta\rho_i-3(\rho_i+p_i)\dot \Psi +(\rho_i+p_i)\vec\nabla\cdot\vec u=0\,,
\end{equation}
where $c_i = \partial p_i/\partial\rho_i$, which is exactly the GR result as expected.

In the next sections we will derive the remaining equations for the two cases of conformally and disformally coupled fluids.

\subsubsection{Conformally coupled fluid}
We now derive the exact expressions for the linear perturbation equations for a conformally coupled perfect fluid. We will first derive the equations for the gravitational potentials $\Phi$ and $\Psi$ and then the one for the spatial components of the fluid velocity.

\paragraph{Euler equation}
\begin{multline}
E_{sc1} \dot u^i+E_{sc2}Hu^i
+E_{sc3}\nabla^i\Phi+3\as E_{sc4}\nabla^i\delta p_c =\\
=\as(\rho_c+p_c)\left[H(\nabla^i\dot\Phi+4\nabla^i\dot \Psi)+3\nabla^i\ddot\Psi+\nabla^i\nabla^2\Phi+2\nabla^i\nabla^2\Phi\right]\,,
\end{multline}
where the background coefficients $E_{sci}$ are reported in appendix \ref{app:linpert}.
Notice that the left hand side of the Euler equation has the same structure of the one derived in GR, only with modified background coefficients. On the contrary the right hand side is represent a characteristic of the NMC. It shows how the velocity is now sourced by a complex combination of derivatives of the metric potentials. \footnote{This is something expected in any NMC model and has been dubbed kinetic mixing. See \cite{Bettoni:2015wta} for an analysis in the context of Horndeski theories.}

\paragraph{Time time component of the Einstein equations} 

\begin{multline}
-2\kappa^2\rho\Phi-6H^2\dot\Psi\left[1+\as\left(2\rho_c+3p_c\right)\right]+2\left(1-\as\rho_c\right)\nabla^2\Psi-\kappa^2\delta\rho =\\
=\as\left[-3H^2\left(2+3c_c^2\right)\delta\rho_c+3 H(\rho_c+p_c)\vec \nabla\cdot \vec u- \nabla^2\delta\rho_c\right]\,.
\end{multline}
The left hand side of this equation is formally identical to the GR expression, only with modified cefficients. The right hand side represent instead a novel feature of the NMC: gradient of the fluid variables are acting as a source for the gravitational potential.

\paragraph{Space space component of the Einstein equations}
\begin{multline}
\as  \nabla_{i}\nabla_{j}\delta \rho_c -(1-\as \rho_c ) (\nabla_{i}\nabla_{j}\Phi -  \nabla_{i}\nabla_{j}\Psi)\bigr)  \\
+ h_{i j} \left[(2 E_{s1} \Phi + 2 E_{s2} \dot \Phi +2 E_{s3} \Psi
 + 6 E_{s4} \dot \Psi + 2 E_{s5} \ddot \Psi  \right.\\
\left.   + (1+\as p_c)\nabla^2\Phi - (1+\as(\rho_c+2p_c)) \nabla^2\Psi
  \right]=\\
  =h_{i j} \left[ 
  +\as \nabla^2\delta \rho_c +\as E_{s7} \delta \rho_c+9 \as H^2 (\rho_c+p_c)\delta c_c^2\right.\\
\left. -2 \as E_{s6} \vec \nabla\cdot \vec u +  \as    (\rho_c + p_c) \vec \nabla\cdot \dot{\vec u}\right]\,,
\end{multline}
where the exact expressions for the background coefficients $E_{si}$ are reported in appendix \ref{app:linpert}.
From the first line is clear that anisotropic stresses are present and are related to gradients of the non-minimal coupling function $F_c=-\rho_c$. This represents a first difference with respect to GR. The part proportional to the spatial projector $h_{\mu\nu}$ contains on the left hand side, the standard GR terms with modified ceofficients, while the right hand side represent the genuine contributions coming from the NMC.

\paragraph{Space time component of the Einstein equations}
\begin{multline}
 \left(\kappa^2  \rho 
 +3\as (2H^2+\dot H)(\rho_c+p_c)\right)u_i  + 2 (H \nabla_{i}\Phi + \nabla_{i}\dot \Psi) \left(1+\frac{\as}{2}(\rho_c+p_c)\right)=\\
= \as\left(  - H (4 - 3 c_c^2) \nabla_i\delta\rho_c
 + (\rho_c + p_c) \nabla_{i}\vec \nabla\cdot \vec u\bigr)\right)\,.
\end{multline}
Also in this case we have on the left hand side the standard GR terms with modified background coefficients while on the right hand side new effect related to the gradients of the fluid variables are present.

\subsubsection{Disformally coupled fluid}

We now move to the analysis of the linear perturbation equations for a disformally coupled perfect fluid. We will first derive the Euler equation and then the equations for the gravitational potentials $\Phi$ and $\Psi$ .

\paragraph{Euler equation}

\begin{multline}
 E_{d1}  \dot u^i+ E_{d2} H u^i
+ E_{d3} \nabla^{i}\Phi  -  3 \ar (H^2 + \dot H) \nabla^{i}\delta p_d=\\
=\ar \left(  H (\rho_d - 3 p_d) \nabla^{i}\dot \Phi -  6  H (\rho_d + 3 p_d) \nabla^{i}\dot \Psi \right.\\
\left. +  (\rho_d - 3 p_d) \nabla^{i}\ddot \Psi -  (\rho_d + p_d) \nabla^{i}\nabla^2\Phi\right)
\end{multline}
where the background coefficients $E_{di}$ are reported in appendix \ref{app:linpert}.

The structure of this equation is similar to the one obtained in the case of conformally coupled fluids. The right hand side contains terms that are also present in the GR case, only with modified background coefficients. On the right hand side instead we find the derivative interaction typical of NMC theories.

\paragraph{Time time component of the Einstein equations}
\begin{multline}
  - 4 \nabla^2\Psi + 2 \kappa^2 (\delta \rho + 2 \rho \Phi) +12 H  \dot \Psi \bigl(1 - \frac{\ar}{2}( \rho_d + 3 p_d)\bigr)=\\
=\ar\left[ H  (\rho_d-3p_d) \vec \nabla\cdot \vec u -
 \nabla^2\delta \rho_d- 3  H^2 \Bigl(1+3c_d^2\Bigr)\delta\rho_d\right]
\end{multline}

\paragraph{Space space component of the Einstein equations}

\begin{multline}
\ar \left(- H (\rho_d + 3 p_d) \frac{\nabla_{i}u_j}{a} + \rho_d \frac{\nabla_{i}\dot u_j}{a} \right)-  \frac{\nabla_{i}\nabla_{j}\Phi -  \nabla_{i}\nabla_{j}\Psi}{a^2} \\
+ h_{i j} \left(E_{sd1}\ddot \Psi  +E_{sd2}  \Phi  + \Psi E_{sd3} + \dot \Psi E_{sd4} + E_{sd5}\dot \Phi + E_{sd6}\frac{\nabla^2\Phi}{a^2} -  \frac{\nabla^2\Psi}{a^2} \right)=\\
=\ar h_{i,j}\left(- E_{sd7} \frac{ \vec\nabla\cdot \vec u }{a}+ \frac{1}{2}  (\rho_d -p_d)\frac{ \vec \nabla \cdot \dot{\vec u}}{ a}+E_{sd8} \delta \rho_d +\frac{9}{2}(\rho_d+p_d)\delta c_d^2\right)
\end{multline}
where the coeffcients $E_{sdi}$ are reported in appendix \ref{app:linpert}.

\paragraph{Space time component of the Einstein equations}

\begin{multline}
 \bigl(2 \kappa^2 (\rho + p) + \ar \dot H (\rho_d - 3 p_d) - 3 \ar H^2 (\rho_d + p_d)\bigr) u_i  \\
+ \bigl(4 - 3 \ar ( \rho_d +  p_d)\bigr)\left( \nabla_ {i}\dot \Psi +  H \nabla_ {i}\Phi \right)=\\
=\ar\left[  \bigl(2 \nabla_i\delta\rho_d + 3\nabla_i\delta p_d\bigr)-  ( \rho_d +  p_d) \nabla_ {i}\vec\nabla\cdot\vec u\right]
\end{multline}

We conclude this section with a comment on the number of degrees of freedom. As it has been discussed in section \ref{sec:dof}, the model under consideration should not propagate more degrees of freedom than those of standard fluid dynamics. Indeed, the investigation of the linear perturbations around a FLRW metric confirms that result: all second time derivatives of the velocity that appears from the various terms in the disformally coupled case exactly cancel each other. Notice that this choice for the metric is much more sensitive to extra degrees of freedom than maximally symmetric space-times (Minkowski or de Sitter). 

\section{Newtonian Limit}
\label{sec:Newton}

To properly discuss the Newtonian limit, it is important to carefully work out the weak field limit of equations \eqref{eq:EFEconf} and \eqref{eq:EFEdisf}.  This was computed in \cite{Bettoni:2011fs}, so here we will report only the main results, referring to interest reader to the cited paper. 

In this limit, the fluid equations of motion for both the conformally and the disformally coupled fluid formally reduce to the standard continuity and Euler equations as it can be easily seen by noting that any combination $R\times F_i\sim (4\pi G)^2$ with $i=c,d$ is a post-Newtonian contribution to the equations. However, the dynamics of the fluid is not equal to that of a minimally coupled fluid as it will receive contributions coming from the modified gravitational potentials. 

The NMC effects remain in the gravitational equation. In particular, the Poisson equation is modified, and the GR equality between the two gravitational potential is broken so that anisotropic stresses are expected. 
As we have said, any combination of the kind $R\times F_i\sim (4\pi G)^2$ is a post-Newtonian contribution to the equations and hence will be dropped. Then we define 
\begin{equation}
g_{\mu\nu} = \eta_{\mu\nu} + \gamma_{\mu\nu} \,, \qquad \bar{\gamma}_{\mu\nu} = \gamma_{\mu\nu} - \frac{1}{2} \eta_{\mu\nu} \gamma\,, \qquad \gamma = \eta^{\mu\nu} \gamma_{\mu\nu}\,,
\end{equation}
so that the modified Einstein equations in the weak field limit in the transverse gauge for the conformally and disformally coupled fluid read respectively
\begin{eqnarray}
-\frac{\mpl}{2} \Box \bar{\gamma}_{\mu\nu}& =& 
-\as \mpl\left(\eta_{\mu\nu} \Box F_c- \partial_\mu\partial_\nu F_c \right) +g_{\mu\nu}\left(F-n\frac{\partial F}{\partial n}\right)-n\frac{\partial F}{\partial n}u_\mu u_\nu 
\,,\\
-\frac{\mpl}{2} \Box \bar{\gamma}_{\mu\nu} &=&  -\ar \frac{\mpl^2}{2} \Omega_{\mu\nu} +g_{\mu\nu}\left(F-n\frac{\partial F}{\partial n}\right)-n\frac{\partial F}{\partial n}u_\mu u_\nu \,,
 \qquad\qquad\qquad\qquad \quad 
\end{eqnarray}
where
\begin{equation}
\Omega_{\mu\nu} =
\delta_{\mu}^{0}\delta_{\nu}^{0}\Box F_d
- \delta_{\nu}^{0} \partial_{0} \partial_{\mu}F_d
- \delta_{\mu}^{0} \partial_{0} \partial_{\nu}F_d
+\eta_{\mu\nu} \partial_{0}\partial_{0}F_d\,. 
\end{equation}

As one immediately sees, the effect of the non-minimal coupling is still present, even in the weak field limit, and the fluid is not behaving as a perfect fluid in Minkowski space-time: the non-minimal coupling has generated a SET which contains additional terms,
constructed out of the derivatives of the fluid variables. 

Putting everything together, and considering the static, non relativistic limit (i.e., the $c^{2} \rightarrow \infty$ limit), we get the Poisson equation for the Newtonian gravitational field in the conformal and disformal case:
\begin{eqnarray}
\nabla^{2} \Phi_{N} &=& 4 \pi G_{N} \left( \rho- \as \nabla^2 \tilde{F}_c \right)\,,\\
\nabla^{2} \Phi_{N} &=& 4 \pi G_{N} \left( \rho- \frac{\ar}{2} \nabla^2 \tilde{F}_d \right)\,,
\end{eqnarray}
where $\tilde F_i =F_i/(4\pi G)$. The Newtonian potential has as sources not only the mass density $\rho$, but also derivative terms. In this sense it will depends not only on how much matter there is at one point but also on how it is distributed.

The spatial part of the Einstein equations give
\begin{eqnarray}
\nabla^2\Psi_{ij}&=&4\pi G\eta_{ij}\left[\rho-\as \nabla^2\tilde F_c\right]+8\pi G\as\left[\eta_{ij}\nabla^2 \tilde F_c+\partial_i\partial_j \tilde F_c\right] \,,\\
\nabla^2\Psi_{ij}&=&4\pi G\eta_{ij}\left[\rho-\frac{\ar}{2}\nabla^2\tilde F_d\right]\,.
\end{eqnarray}
Making use of the Poisson equation the two spatial equations can be rewritten as
\begin{eqnarray}
\nabla^2\Psi_{ij}&=&\eta_{ij}\nabla^2\Phi +8\pi G\as\left[\eta_{ij}\nabla^2\tilde F_c+\partial_i\partial_j\tilde F_c\right] \,,\\
\nabla^2\Psi_{ij}&=&\eta_{ij}\nabla^2\Phi\,,
\end{eqnarray}
which interestingly shows how, for a disformal coupling, there is only one gravitational potential, while the conformal coupling can be responsible for anisotropic stresses. 

To summarize, we have seen how both forms of NMC produce a derivative correction to the Poisson equation but while conformally couple fluids will excite extra gravitational degrees of freedom with respect to GR and will have anisotropic stresses, a disformally coupled fluid is described by a single gravitational potential and has no anisotropic stresses.

A comment is also in order about the presence of the Laplacian in the Poisson equations. Indeed, this implies an algebraic relation between the gravitational potential and the NMC functions which are ultimately related to the matter distribution. As it was noted in~\cite{Sotiriou:2008dh}, this implies that any sharp change in the matter distribution can be imprinted in an analogous change in the gravitational potential thus being potentially dangerous for the kind of models under investigations. However, in our case, the NMC matter has a much shallower distribution as compared to standard baryonic matter investigated in the cited paper, being related to the description of the dark component of the Universe. Also, sufficiently smooth functions $F_c$ or $F_d$ can avoid such issues. Hence, we do not expect that this feature of the modified Poisson equations will practically affect the physical viability of this class of models.

Finally, it is perhaps worth noticing that that, when considering both conformal and disformal couplings, precise cancellations can happen leading to a dynamics which would be very similar to the GR one.
For example one might consider the case in which the two contributions to the Poisson equation  from these couplings cancel each other thus leaving the modifications of gravity only to the spatial part of the potentials. Indeed this happens for the particular combination
\begin{equation}
\as=-\frac{\ar}{2}\qquad \textrm{and} \qquad F_c=F_d\,,
\end{equation}
which corresponds to the following NMC
\begin{equation}
\mathcal{L}_{\textrm{NMC}}=\sqrt{-g}\Bigl[\alpha_{RG} F_{RG}\left(-R+G_{\mu\nu}u^\mu u^\nu\right)\Bigr]\,,
\end{equation}
which, remarkably, involves exactly a coupling of the Horndeski type via the Einstein tensor.

\section{Dark Matter-Dark Energy non-minimal couplings?}
\label{sec:DE_DM}

In this section we will discuss another intriguing possibility of NMC fluids. In fact, one can speculate whether there could be more than one fluid which is NMC. In particular, one can imagine the situation in which there are two NMC fluids one playing the role of DM and another one related to DE, such that couplings of the form
\begin{equation}
F_1 (n_1,s_1)u_1^\mu R_{\mu\nu}u_1 ^\nu+F_2(n_2,s_2) u_2^\mu R_{\mu\nu}u_2^\nu +F_{12}(n_1,s_1;n_2,s_2) u_1^\mu R_{\mu\nu}u_2^\nu\,,
\end{equation}
can be available. The last term in particular, represents an interaction, mediated by gravity, between DE and DM fluids. Recently, it has been shown that Planck data allow for this possibility \cite{Pettorino:2013oxa} or even more, they seems to favour such interaction \cite{Salvatelli:2014zta,Abdalla:2014cla} so that coupling as those presented above may represent an interesting extension to the model presented in this paper able to include such experimental hints.
Interactions of this kind are also interesting in connection to other models like the one investigated in \cite{Blas:2012vn}, where a Lorentz breaking vector field is coupled to a fluid DM fluid, or the one investigated in \cite{Jimenez:2013qsa} where a vector version of the Horndeski action is constructed.

If we consider the simplest model in which only the NMC between the two fluids is present we get the Einstein equations
\begin{multline}
 M_{\text{Pl}}^2G_{\mu\nu} -T_{\mu\nu}^{(1)}-T_{\mu\nu}^{(2)}
 + \alpha_R \left[ -\nabla_\alpha\nabla_{(\mu} t^{(12)}_{\nu)}{}^{\alpha} + \Box t^{(12)}_{\mu\nu} + g_{\mu\nu} \nabla_\alpha\nabla_\beta t^{(12)\alpha\beta}\right.  \\
\left.+\enangle{R}\left(F_d  g_{\mu\nu} + h^{(1)}_{\mu\nu} \left(n_1 \frac{\partial F_d}{\partial n_1}-F_d\right) + h^{(2)}_{\mu\nu} \left(n_2 \frac{\partial F_d}{\partial n_2}-F_d\right)\right)\right]=0\,,
\label{eq:EFE_DM-DEint}
\end{multline}
where $T_{\mu\nu}^{(i)}$ is the SET for the i-th component as given by \eqref{SET_F}, $h^{(i)}_{\mu\nu}=g_{\mu\nu}+u^i_\mu u^i_\nu$, while $t^{(12)}_{\mu\nu} = (F_d u^1_{(\mu} u^2_{\nu)})/2$.
The fluid are better derived from the direct variation of the action with respect to the fluid variables of both components.
In particular, unless some mixing between the constraint is introduced, both entropies and number densities will be separately conserved. On the other hand, the Euler equation gets modified as
\begin{multline}
\left(\frac{\partial F_i}{\partial n_i} -\alpha_{12}\enangle{R}\left(\frac{\partial F_{12}}{\partial n_i}-\frac{F_{12}}{n_i}\right)\right)\dot{u}^\alpha + h_{(i)}^{\alpha\rho}\nabla_\rho\left[\frac{\partial F_i}{\partial n_i}-\alpha_{12}\enangle{R}\left(\frac{\partial F_{12}}{\partial n_i}-\frac{F_{12}}{n_i}\right)\right] =\\
=-\alpha_{12} h_{(i)}^{\alpha\nu}u_i^\sigma\left[\nabla_\nu\left(\frac{F_{12}}{n_i}R_{\mu\sigma}u_j^\mu\right)-\nabla_\sigma\left(\frac{F_{12}}{n_i}R_{\mu\nu}u_j^\mu\right)\right]\,,
\end{multline}
where $i,j=1,2$, $i\neq j$ and $F_i=F_i(n_i)$ and $F_{12}=F_{12}(n_1,n_2)$.

To summarize, in the case of two interacting fluids, we end up with a system in which the two fluids have conserved charges along the flow, $n_i$ and $s_i$ and a modified Euler equations. In particular, the inhomogeneities in each of the fluid act as force terms for the other.

\section{Discussion and conclusions}
\label{sec:Conclusions}
%
%
%
%
%

In the era of precision cosmology and with many upcoming cosmological and astrophysical surveys, alternatives to the \lcdm will be dramatically put under scrutiny and most likely we will soon be able to distinguish among the different proposals. 
In this spirit and motivated by the quest for a better understanding of the dark component of the Universe and of the fluid dynamics in curved space-times, we have investigated the full dynamics of a NMC fluid where both a conformal and a disformal coupling to curvature are present. We have seen how the requirement of no fluid derivative in the action reduces the possible NMC only to two terms, $F_c R$ and $F_d R_{\mu\nu} u^\mu u^\nu$. These two terms are however general enough to produce significant modifications with respect to the simple minimally coupled case.

In particular, the Einstein equations are much richer and shows the appearance of derivative terms of fluid variables which play an important role in the Newtonian limit. In fact, the Poisson equation may be sourced by gradients of the fluid density distribution meaning that in these scenarios it will depend not only on how much matter is there at a point but also on how it is distributed. We have also pointed out how the conformal and the disformal couplings have different features: while the former is related to anisotropic stresses the latter is not, thus it does not excite new gravitational degrees of freedom with respect to GR. These features directly relate to observable quantities and hence are potential tools for constraining this class of models.

For what concerns the fluid equations, we show how in the Euler equation appears an extra force term related to the curvature, which is a clear and distinguishable feature of the NMC, while the continuity equation is modified in such a way that it will still allow the definition of conserved quantities along the flow.  
Hence, structure formation will be affected by the NMC in a way that may potentially stabilize their growth under a certain scale (as a consequence of the extra force). On the other hand, as one can see from the Friedmann equations, changes are expected to occur also at the background level. The model can be easily made compatible with early universe physics, which is well described by the \lcdm model, while still providing interesting new features at present.

We have also discussed the important issue of extra degrees of freedom that may be propagated in this kind of models as a consequence of the NMC. In section \ref{sec:dof} we have provided an argument for why this should not be the case. Of course, only a complex Hamiltonian analysis can confirm this conclusion but we reserve this analysis for a future work. It is also important to stress that the model presented in this paper is a phenomenological one and not a fundamental theory of gravity and that it has been though as a cosmological extension to the \lcdm model. For this reason, in section \ref{sec:linpert}, we have investigated the dynamics of linear perturbations around a cosmological background and showed how it is free of extra degrees of freedom. We hence conclude that in the regime for which the model has been thought to apply to there are no unwanted degrees of freedom.

Finally we have considered the case of two NMC fluids which can interact via curvature couplings. This is an intriguing possibility as it may allow DM-DE interactions an interesting extension that seems to be favoured by current cosmological data.

The issue of non-minimal couplings in gravitation theories is a long standing one. When dealing with cosmological fluids we have argued that non-minimal coupling for exotic components like DM and DE should not and cannot be neglected (even just to be able to exclude them on the base of future observations). We have presented here an exploration of the implications of this extension of the standard cosmological framework and found several interesting novelties that we think should stimulate deeper investigations of these models.  We think that NMC fluid models are now mature for phenomenological applications and we hope to further advance in this direction in future.

\acknowledgments

The authors wish to thank Lorenzo Sindoni for his careful reading of a previous version of the draft and the referee for the useful comments that helped improving the work. D. B. would like to thank Jose Beltran Jimenez and Adam Solomon for the discussions had while staying at NORDITA. D. B. acknowledges support from the I-CORE Program of the Planning and Budgeting Committee, THE ISRAEL SCIENCE FOUNDATION (grants No. 1829/12 and No. 203/09), and the Asher Space Research Institute. S. L. acknowledges financial support from the John Templeton Foundation (JTF), Grant No. 51876.
The calculations have been checked with the package xAct \cite{Brizuela:2008ra,xAct}.

\appendix

\section{Equation of motion from the fluid variables}

In this appendix we report the equations derived from the NMC actions \eqref{eq:CCFA} and \eqref{eq:DCFA} when the variation is taken with respect to the fluid variables. This has been used as a consistency check that the equations obtained from the covariant conservation of the metric equation and those derived from a direct variation of the action with respect to the fluid variables are equivalent. Moreover, the equations presented in this appendix help in clarifying some thermodynamic aspects of the NMC fluid, in particular making explicit the way they modify the standard relations.

\subsection{The case of conformally coupled fluid}
\label{app:FEOM}

The variation of the action \eqref{eq:CCFA} with respect to the fluid variables gives the following set of equations
\begin{eqnarray}
0 &=& \frac{\delta S}{\delta J^\mu}=-u_\mu\left(\frac{\partial F}{\partial n}+\as R \frac{\partial F_c}{\partial n}\right)+\nabla_\mu\varphi +s\nabla_\mu\theta +\beta_A\nabla_\mu\alpha^A\,,\label{eq:constraints_CCF_Euler}\\
0 &=& \frac{\delta S}{\delta \varphi}=-\nabla_\mu J^\mu\,,\label{eq:constraints_CCF_number}\\
0 &=& \frac{\delta S}{\delta \theta}=-\nabla_\mu(sJ^\mu)\,,\label{eq:constraints_CCF_entropy}\\
0 &=& \frac{\delta S}{\delta s}=\sqrt{-g}\left(\frac{\partial F}{\partial s}+\as R\frac{\partial F_c}{\partial s}\right)+J^\mu\nabla_\mu\theta\,,\label{eq:constraints_CCF_temperature}\\
0 &=& \frac{\delta S}{\delta \alpha^A}=-\nabla_\mu(\beta_AJ^\mu)\,,\label{eq:constraints_CCF_lagrangian1}\\
0 &=& \frac{\delta S}{\delta \beta_A}=J^\mu\nabla_\mu\alpha^A\,,
\label{eq:constraints_CCF_lagrangian2}
\end{eqnarray}
Only two equations are modified with respect to those obtained from a minimally coupled fluid \eqref{eq:EulerPF1}-\eqref{eq:constraints} and as it has been discussed the modifications occurs as rescalings or shifts of existing fluid variables. For example, the first equation, contracted with the four velocity $u^\mu$, and the fourth one, reads
\begin{eqnarray}
-\frac{\partial}{\partial n}\left(F+\as R F_c\right)&=&\dot\varphi +s\dot \theta\,,\\
-\frac{\partial}{\partial s}\left(F+\as R F_c\right)&=& \dot{\theta}\,.
\end{eqnarray}
Given the similarity if the right hand side of these equations with those for the minimally coupled case it is straightforward to make the following thermodynamic identifications
\begin{eqnarray}
\rho_c &=&-(F+\as R F_c)\,,\\
\dot{\varphi}&=& f\,,\\
\dot{\theta} &=& T\,.
\end{eqnarray}
where $T$ is the temperature and  $f$ is the chemical free energy, as defined in section \ref{sec:PFA}.

From the set of equations \eqref{eq:constraints_CCF_Euler}-\eqref{eq:constraints_CCF_lagrangian2} it is easy to show that one gets the same equations as from the conservation of the Einstein equations. In fact, from equations \eqref{eq:constraints_CCF_number} and \eqref{eq:constraints_CCF_entropy} we recover equation \eqref{eq:continuity_C}, while from equation \eqref{eq:constraints_CCF_Euler}, using the procedure described at the end of section \ref{sec:PFA}, we recover exactly equation \eqref{eq:Euler_CCF}.
Hence, we conclude that the addition of the conformal NMC is compatible with the assumption to have a perfect fluid. 

\subsection{The case of disformally coupled fluid}
We proceed now with the computation of the disformally coupled equations from the action \eqref{eq:DCFA}. The variation with respect to the fluid variables gives the following system of equations is
\begin{eqnarray}
\nonumber
0 &=& \frac{\delta S}{\delta J^\sigma}=-\frac{\partial F}{\partial n} u_\sigma +\nabla_\sigma\varphi +s\nabla_\sigma\theta +\beta_A\nabla_\sigma\alpha^A-\frac{2}{n}\ar\left(n\frac{\partial F_d}{\partial n}\enangle{R} u_\sigma-F_d R_{\mu\nu}u^\nu h^\mu{}_\sigma\right)\,,\\
\label{eq:constraints_DCF_Euler}\\
0 &=& \frac{\delta S}{\delta \varphi}=-\nabla_\mu J^\mu\,,\label{eq:constraints_DCF_number}\\
0 &=& \frac{\delta S}{\delta \theta}=-\nabla_\mu(sJ^\mu)\,,\label{eq:constraints_DCF_entropy}\\
0 &=& \frac{\delta S}{\delta s}=\sqrt{-g}\left(\frac{\partial F}{\partial s}+2\ar \enangle{R}\frac{\partial F_d}{\partial s}\right)+J^\mu\nabla_\mu\theta\,,\label{eq:constraints_DCF_temperature}\\
0 &=& \frac{\delta S}{\delta \alpha^A}=-\nabla_\mu(\beta_AJ^\mu)\,,\label{eq:constraints_DCF_lagrangian1}\\
0 &=& \frac{\delta S}{\delta \beta_A}=J^\mu\nabla_\mu\alpha^A\,,
\label{eq:constraints_DCF_lagrangian2}
\end{eqnarray}
where as for the previous case of conformally couple fluid only two equations are modified. These modifications occurs in a very similar way as what happens in the case of the conformally coupled fluids, the only difference being the last term in the first equations. However, when contracted with the four velocity of fluid such term disappears. Hence, we can define as in the previous case the following thermodynamic quantities
\begin{eqnarray}
\rho_d &=&-(F+2\ar \enangle{R} F_d)\,,\\
\dot{\varphi}&=& f\,,\\
\dot{\theta} &=& T\,.
\end{eqnarray}
where again $T$ is the temperature and  $f$ is the chemical free energy, as defined in section \ref{sec:PFA}.

Also, using the same procedure described above one can show that the equations for the fluid \eqref{eq:constraints_DCF_Euler}-\eqref{eq:constraints_DCF_lagrangian2} implies those obtained  form the conservation of the Einstein equations.

\section{Linear perturbation coefficients}
\label{app:linpert}

In this appendix we report the detailed expressions of the linear perturnation coefficients used in section \ref{sec:linpert}.

\subsection{Conformally coupled fluid}

The coefficients of the space space component of the Einstein equations are
\begin{eqnarray}
E_{s1} & = & (3 H^2 + 2 \dot H) +  \as \Bigl(2 \dot H (2 \rho_c + 3 p_c) + H^2 \bigl(3 p_c - 9 c_c^2 (\rho_c + p_c)\bigr)\Bigr)\,,\\
E_{s2} & = &  H +  \as  H (2 \rho_c + 3 p_c)\,,\\
E_{s3} & = &  (3 H^2 + 2 \dot H + \kappa^2 p) +  \as  \Bigl(2 \dot H (2 \rho_c + 3 p_c) + H^2 \bigl(3 p_c - 9 c_c^2 (\rho_c + p_c)\bigr)\Bigr)\,,\\
E_{s4} & = &  H -  \as H \bigl(- p_c + 3 c_c^2 (\rho_c + p_c)\bigr)\,,\\
E_{s5} & = & 1 +  \as (2 \rho_c + 3 p_c)\,,\\
E_{s6} & = & \bigl(3 H - (1 - 3 c_c^2) H\bigr) (\rho_c + p_c)\,,\\
E_{s7} & = & H^2(2-10c_c^2 +9c_c^4)+3\dot H (1+2c_c^2)\,.
\end{eqnarray}

The coefficients in the Euler equation are
\begin{eqnarray}
E_{sc1} & = & \rho+3\as(\rho_c+p_c)(2H^2+\dot H)\,,\\
E_{sc2} & = &\rho+3\as(\rho_c+p_c)\left(2H^2(1-3c_c^2)+\dot H(5-3c_c^2)+\frac{\ddot H}{H}\right)\,,\\
E_{sc3} & =&  \rho-3\as(2H^2+\dot H)(\rho_c+p_c)\,,\\
E_{sc4} & = & (2H^2+\dot H)\,.
\end{eqnarray}

\subsection{Disformally coupled fluid}

The coefficients for the Euler equation read
\begin{eqnarray}
E_{d1} & = &  \rho +  p - \ar \bigl(- \dot H (\rho_d - 3 p_d) + 3 H^2 (\rho_d + p_d)\bigr)\,,\\
E_{d2} & = & (1 - 3 c_s^2)  (\rho + p) - \ar \Bigl(- \frac{\ddot H}{H} (\rho_d - 3 p_d) \\
&-& 3 (-1 + 3 c_d^2) H^2 (\rho_d + p_d) 
 +   \dot H \bigl((5 - 9 c_d^2) \rho_d + 3 (7 - 3 c_d^2) p_d\bigr)\Bigr)\,,\\
 E_{d3} & = &  \rho + p - \ar \bigl(\dot H (\rho_d - 3 p_d) - 3 H^2 (\rho_d + 5 p_d)\bigr)
\end{eqnarray}

Coeffcienents of the space space component of the Einstein equations
\begin{eqnarray}
E_{sd1} & = & 2 - \ar \bigl(\rho_d + 3 p_d \bigr)\,,\\
E_{sd2} & = & 6 H^2 + 4 \dot H + \ar \Bigl(3 H^2 \bigl(3c_d^2\rho_d -(2-3c_d^2)p_d\bigr)  -  2 \dot H \bigl(\rho_d+3p_d\bigr)\Bigr)\,,\\
\nonumber
E_{sd3} & = & E_{ds2} 
\,,\\
 E_{sd4} & = & 6 H - 3 \ar H \bigl(2p_d-3(\rho_d+p_d)c_d^2\bigr)\,,\\
 E_{sd5} & = & 2 H -  \ar H (\rho_d + 3 p_d)\,,\\
 E_{sd6} & = & 1 -  \frac{\ar}{2 } (\rho_d + p_d)\,,\\
 E_{sd7} & = & H\bigl(2p_d-3(1-c_d^2)(\rho_d+p_d)\bigr)\,,\\
 E_{sd8} & = & -\frac{3}{2}H^2\left(4+3c_d^2\right)c_d^2+\dot H \left(1+3 c_d^2)\right)
\end{eqnarray}

\bibliographystyle{JHEP}
\bibliography{mybib}

\end{document}